\documentstyle[seceq,epsf]{ptptex}

\title{
Numerical Study of Aging Phenomena \\
in Short-Ranged Spin Glasses
}

\author{
Koji {\sc Hukushima}, 
Hajime {\sc Yoshino}
and Hajime {\sc Takayama}
}

\inst{%
Institute for Solid State Physics, Univ. of Tokyo, 7-22-1
  Roppongi, Minato-ku, Tokyo 106-8666,  Japan}

\recdate{
}

\abst{%
Aging phenomena of short-range Ising spin glass models have been
investigated using Monte Carlo simulations. 
It is found that in the low-temperature spin-glass phase the mean
domain size exhibits a crossover from a power-law growth associated
with the critical fluctuation at the transition temperature to slower
growth inherent in the low-temperature phase. 
The temperature dependence of the growth law of the domain size can be
almost explained by this crossover. 
We also find that the spin-autocorrelation function in the
quasi-equilibrium regime follows the expected scaling behavior from the
droplet theory
expressed in terms of the mean domain size and  a characteristic length
scale of droplet excitations.  
}
\begin{document}

\maketitle

\section{Introduction}

Spin glasses (SG) exhibit characteristic slow dynamics below the SG
transition temperature.\cite{Review1,Review2,Review3}
Recently such slow dynamics, in particular aging phenomena,
has been much attractive both in experimental and theoretical studies.
Several attempts have been made so far to explain the slow dynamics. 
There are mainly two distinct phenomenological approaches;
one is a phase-space approach, in which the dynamics is described by  
a diffusion in the hierarchically constructed phase space, inspired by
the mean-field theory which suggests multi-valley structure of the phase
space. 
The other is a real-space picture based on the scaling theory\cite{FH1}
in which   
low-lying excitations are attributed to connected clusters reversed from
one of two ground states.
There has been, however, no satisfactory description from a microscopic
model, except for the dynamical mean-field theory.\cite{CK}  
In this paper we present the results on the non-equilibrium dynamics
obtained by large-scale Monte Carlo (MC) simulations on short-range 
Edwards-Anderson (EA) Ising SG models. 
Our analyses of the obtained data are based on the droplet theory. 
We believe that,  even when the phase-space approach may give us a
correct description for the slow dynamics, what really occurs in the
real space is also indispensable for thorough understandings of the
aging phenomena.  

Let us explain briefly the droplet theory.\cite{FH1}  
According to the theory, 
aging process is described by coarsening of domain walls, which is driven
 by successive flipping of thermally activated droplets. 
During isothermal aging up to waiting time $t_w$ after quench,   
domains with the mean size $R(t_w)$ separating different pure states
 have grown up. 
Within each domain, small droplets of size $L\sim L(\tau)\ll R(t_w)$ are
thermally  fluctuating within a time scale of $\tau$ as in
equilibrium. The typical value of their excitation gap $F_L^{\rm
typ}$ scales as  
\begin{equation}
 F_L^{\rm typ}\sim \Upsilon (L/L_0)^\theta, 
\label{eqn:fgap}
\end{equation}
where $\Upsilon$ is the stiffness constant and $L_0$ is a microscopic
 length scale, and that of free-energy barrier $B_L^{\rm typ}$ also
 scales as 
\begin{equation}
 B_L^{\rm typ}\sim \Delta (L/L_0)^\psi,
\end{equation}
where $\Delta$ is a characteristic free-energy scale. 

As compared with the equilibrium, some droplets which touch the  domain
 wall with the length scale $R(t_w)$ could reduce their excitation gap
 from (\ref{eqn:fgap}).  
This effect is estimated by the droplet theory as the reduction of the
 averaged excitation gap which is given by 
\begin{equation}
 F_{L,R}^{\rm typ} = \Upsilon_{\rm eff}(L/L_0)^\theta, 
\end{equation}
with the effective stiffness constant 
\begin{equation}
 \Upsilon_{\rm eff} = \Upsilon \left(1-c_v(L/R)^{d-\theta}\right), 
\label{eqn:effU}
\end{equation}
where $c_v$ is a numerical constant. 
Physical quantities such as the spin autocorrelation function are 
estimated in terms of such length scales 
by taking into account statistical weights of the droplet
excitations appropriately. 
The growth law of the length scales of the domain $R(t_w)$ and the droplet
$L(t)$ is given by 
\begin{equation}
 R(t), L(t) \sim \left(\frac{T}{\Delta}\ln(t/\tau_0)\right)^{1/\psi}, 
\label{eqn:glaw}
\end{equation}
where $\tau_0$ is a microscopic time scale.
One notices that the droplet theory for aging phenomena consists of 
two almost independent steps;
the scaling argument based on the typical length scales of $R(t_w)$ and
$L(\tau)$ and the growth law of these length scales. 
The main purpose of the present work is to test these two steps
separately.

The present paper is organized as follows:
 in the next section after introducing the model system studied, time
evolution of the length scale of domain wall is discussed. 
The results of spin-autocorrelation function in the quasi-equilibrium regime 
of isothermal aging are presented in Sect. 4.

\section{Model and Method}
We focus on the four-dimensional (4D) Ising SG model, 
because its static critical properties have been established  in
the sense that a SG phase transition occurs at finite temperature with a
rigid order parameter.  
The 4D EA Ising SG model is defined by Hamiltonian, 
\begin{equation}
 {\cal H} = -\sum_{\langle ij\rangle}J_{ij}S_iS_j,
\end{equation}
where the sum runs over nearest-neighbor sites and the Ising variables
$S_i$ are defined on a hypercubic lattice with periodic boundary
conditions. 
The interactions are bimodal variables ($J_{ij}=\pm 1$) distributed
randomly with equal probability.  
The simulation method is the standard single-spin-flip
Monte Carlo (MC) method using two-sublattice dynamics with the heat-bath 
transition probability. 
Using the MC method,  we have simulated aging
processes after a rapid quench from $T=\infty$ to the SG phase. 
The system size studied is mainly $L=24$, while the size $L=32$ is
partly studied. We have found no significant difference in the data of
$L=24$ and $32$.

\section{Growth Law of Domain Size}

In order to extract a length scale characterizing the growth of
ordering, we calculate spatial replica correlation function in
off-equilibrium, defined as 
\begin{equation}
 G(r,t) = \sum_i\langle S_i^{(\alpha)}(t)S_i^{(\beta)}(t)
  S_{i+r}^{(\alpha)}(t)S_{i+r}^{(\beta)}(t)\rangle, 
\end{equation}
where $\alpha$ and $\beta$ denote the replica indices which are updated
independently and with different initial spin configurations. The bracket
$\langle\cdots\rangle$ denotes the average over independent bond
realizations. In our simulations, only one MC sequence is performed for
each random bond configuration. 
We extract the mean domain size $R(t)$ by directly fitting $G(r,t)$ to
an exponential form.  

\begin{figure}[htb]
 \parbox{\halftext}{
 \epsfxsize=\halftext
 \centerline{\epsfbox{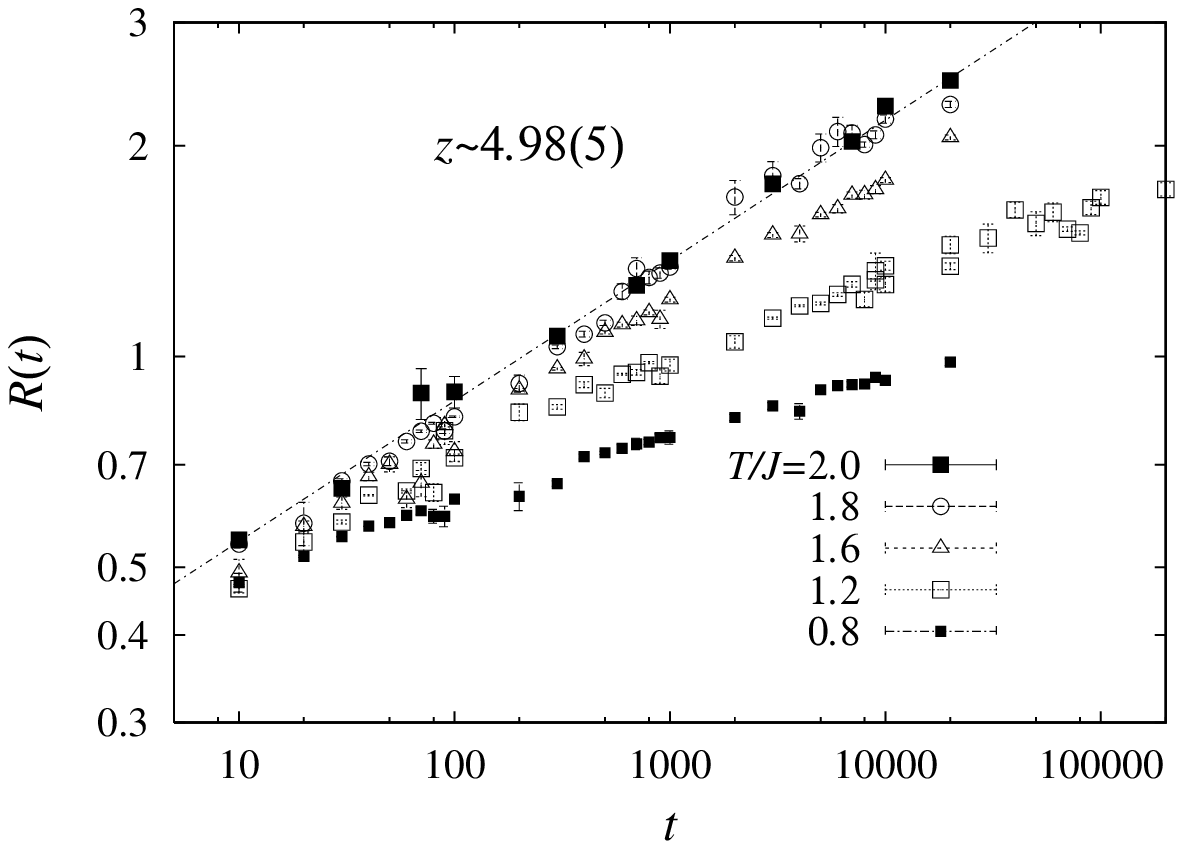}}
 \caption{$R(t)$ of  the 4D Ising SG model  as a function of $t$ at
 $T_{\rm c}$ and below  $T_{\rm c}$.  
}
 \label{fig:xi}}
\hspace{8mm}
\parbox{\halftext}{
 \epsfxsize=\halftext
 \centerline{\epsfbox{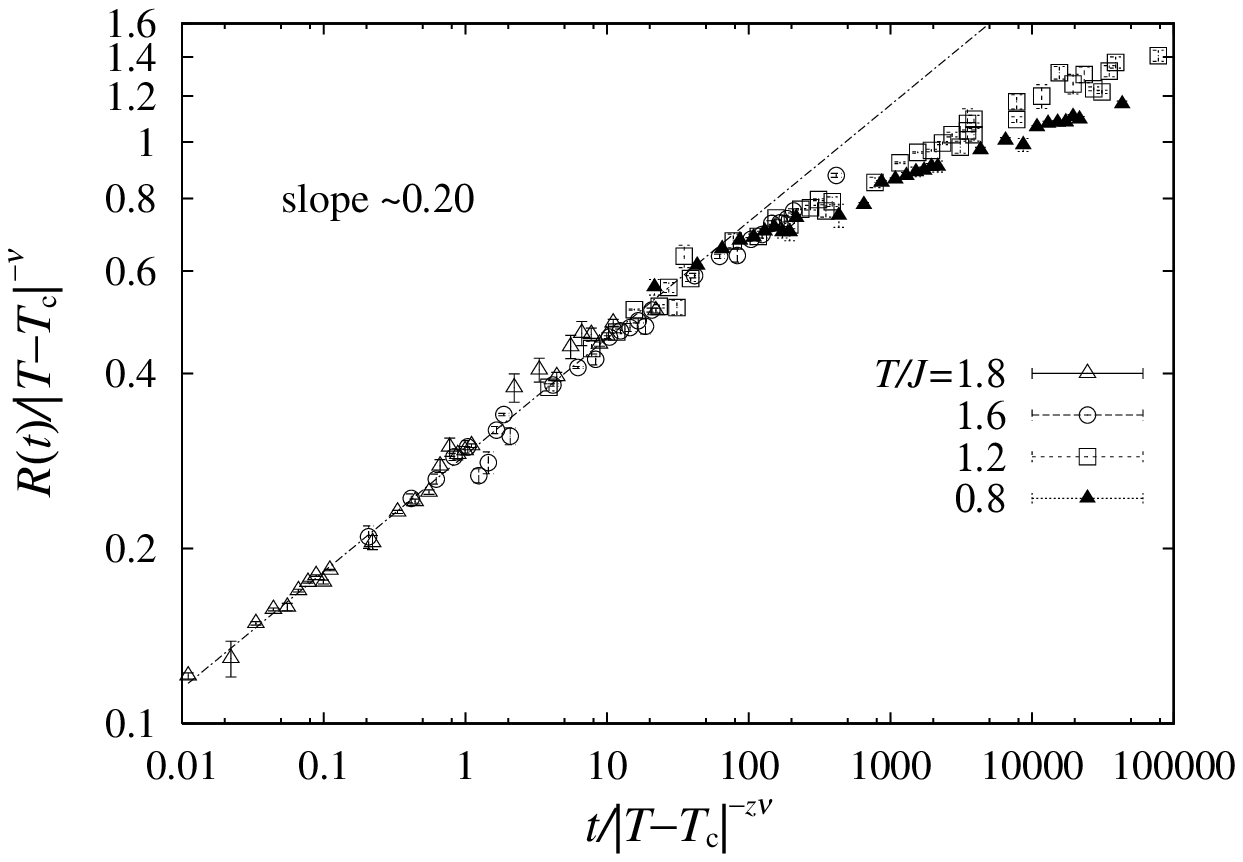}}
 \caption{Scaling plot of $R(t)$ of  the 4D Ising SG model, where the SG 
 transition temperature $T_{\rm c}=2.0J$ and the critical exponent $\nu=0.93$ 
 are fixed.
}
 \label{fig:xi-scal}}
\end{figure}

In Fig.~\ref{fig:xi} we show time dependence of $R(t)$ at the SG
transition temperature $T_{\rm c} (\simeq 2.0J)$ and below. 
Just at $T_{\rm c}$, the length scale is expected to grow as a power law
with the dynamical critical exponent $z$, $R(t)\sim t^{1/z}$, 
irrespective of the physical picture underlying the ordered phase.  
As expected, it is found that  the length scale follows 
a power law.  The exponent $z$ is estimated to be $4.98(5)$, which is
consistent with that of the previous work.\cite{BernardiCampbell}   
Below $T_{\rm c}$, $R(t)$ grows with time slower and slower 
as temperature decreases. 
We try to see crossover between the critical fluctuation and slow 
dynamics inherent in the low-temperature phase. 
In off-equilibrium both at the critical and the off-critical 
temperatures, the length scale would exhibit 
a power law in short length and time regime  where the critical
fluctuation dominates the dynamics.  
We assume that such microscopic length $R_0$ and time $\tau_0$ are
the correlation length and time associated with critical fluctuation in
equilibrium, respectively.   
Thus, we propose a scaling form
\begin{equation}
R(t)/R_0 = g(t/\tau_0),
\end{equation}
with $R_0 =  |T-T_{\rm c}|^{-\nu}$ and $\tau_0 = |T-T_{\rm c}|^{-z\nu}$. 
The scaling plot of $R(t)$ is shown in Fig.~\ref{fig:xi-scal}. 
As expected from a standard scaling theory of critical phenomena, 
the scaling function $g(x)$ for smaller $x$ exhibits a power law
$x^{1/z}$ associated with the critical temperature. 
We find a significant deviation from the power law at longer times, 
suggesting that 
the characteristic slow dynamics of the SG phase takes place there. 
In fact, the functional form is not incompatible with a power law
of $\ln (t)$ predicted by the droplet theory (\ref{eqn:glaw}). 
It is noted that the strong temperature dependence of $R(t)$ shown in 
Fig.~\ref{fig:xi} can be almost explained by introducing the microscopic
units $R_0$ and $\tau_0$ associated with the critical fluctuation, 
while it is not sure whether the asymptotic form at longer times is also
described by a universal scaling function or not.

\section{Scaling  analysis in quasi-equilibrium}

Recent studies have revealed that the off-equilibrium dynamics in the SG
is separated into two characteristic time regimes. One is a short-time
regime, called ``quasi-equilibrium regime'', and the other is a
long-time regime, called ``aging regime''. 
A typical observable to see such two time regimes is the spin
auto-correlation function 
\begin{equation}
 C(\tau ; t_{w}) = \frac{1}{N}\sum_i\langle S_i(t_{w})S_i(\tau+t_w)\rangle,
\end{equation}
where $t_w$ denotes a waiting time after the rapid quench. 
Our interest is in its behavior in the quasi-equilibrium regime, namely
$t_w\gg\tau$. 
Based on the droplet argument using the effective stiffness constant
(\ref{eqn:effU}), the behavior of $C(\tau ; t_w)$  is 
explicitly given by\cite{KYT2} 
\begin{equation}
C(\tau;t_{\rm w})  = C_{\rm eq}(\tau) + 
  \frac{c}{\Upsilon}\frac{T}{(L(\tau)/L_0)^{\theta}} 
\left(\frac{L(\tau)}{R(t_{\rm w})}\right)^{d-\theta} 
 +\cdots , 
\label{eqn:coft}
\end{equation}
with the equilibrium part $C_{\rm eq}(\tau)$\cite{FH1}
\begin{equation}
 C_{\rm eq}(\tau) = q_{\rm EA} + \frac{A}{(L(\tau)/L_0)^\theta}, 
\label{eqn:ceq}
\end{equation}
where $q_{\rm EA}$ is the EA order parameter and $c$ and $A$ are
numerical constants.  
This gives us an extrapolation form of the large $\tau$ limit, namely 
a way of determining the EA order parameter. 
An empirical form $C(t)=q_{\rm EA} + a/t^\alpha$ has been  used
frequently for estimating $q_{\rm EA}$ from the spin autocorrelation
function.\cite{Parisi}  This is true only if the time dependence of 
length scale 
$L(t)$ in (\ref{eqn:ceq}) is  a power law.  
As seen in the last section, however, the observed length scale $R(t)$ 
in this model exhibits the crossover from the critical power law to
slower growth at large $t$. 

\begin{figure}[htb]
 \parbox{\halftext}{
 \epsfxsize=\halftext
 \centerline{\epsfbox{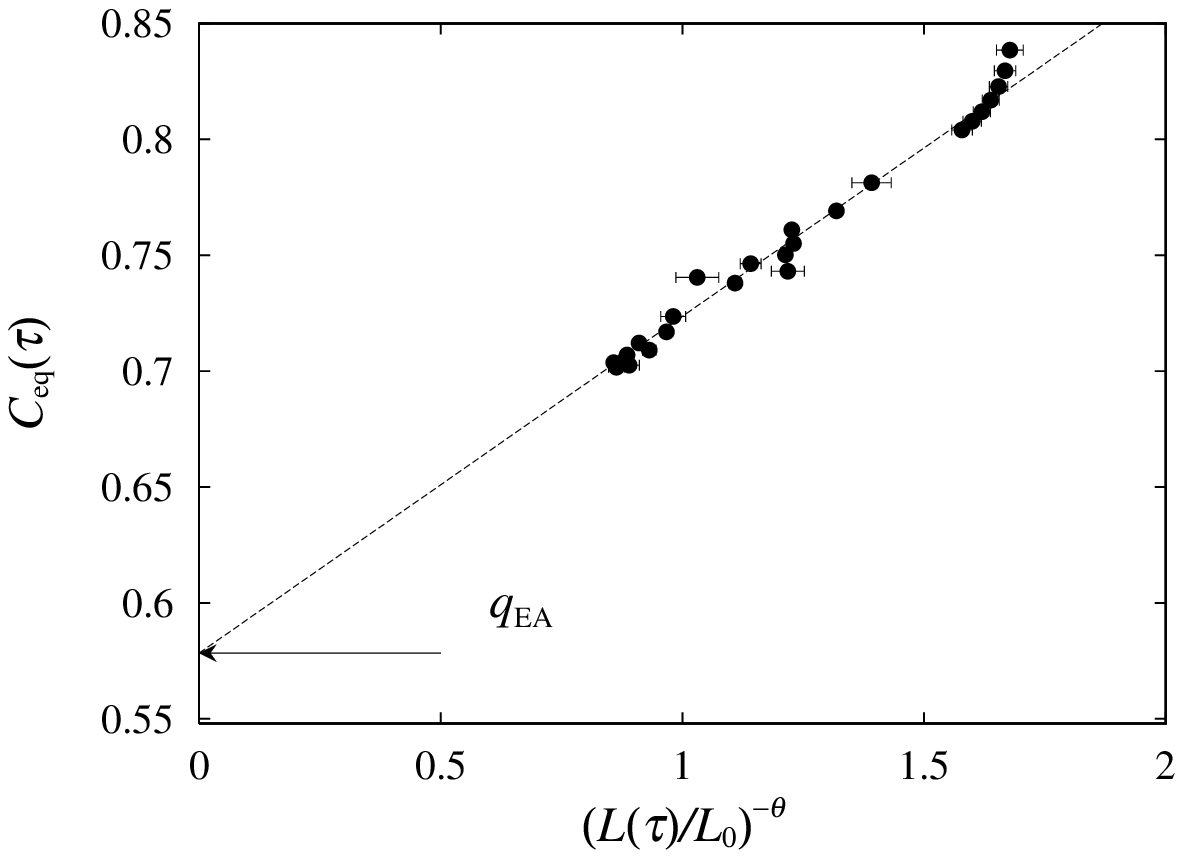}}
 \caption{Equilibrium autocorrelation function extracted to the large
 $t_w$ limit. The line represents a fitting according to the expression
 (\ref{eqn:ceq}).  
}
 \label{fig:ceq}}
\hspace{8mm}
\parbox{\halftext}{
 \epsfxsize=\halftext
 \centerline{\epsfbox{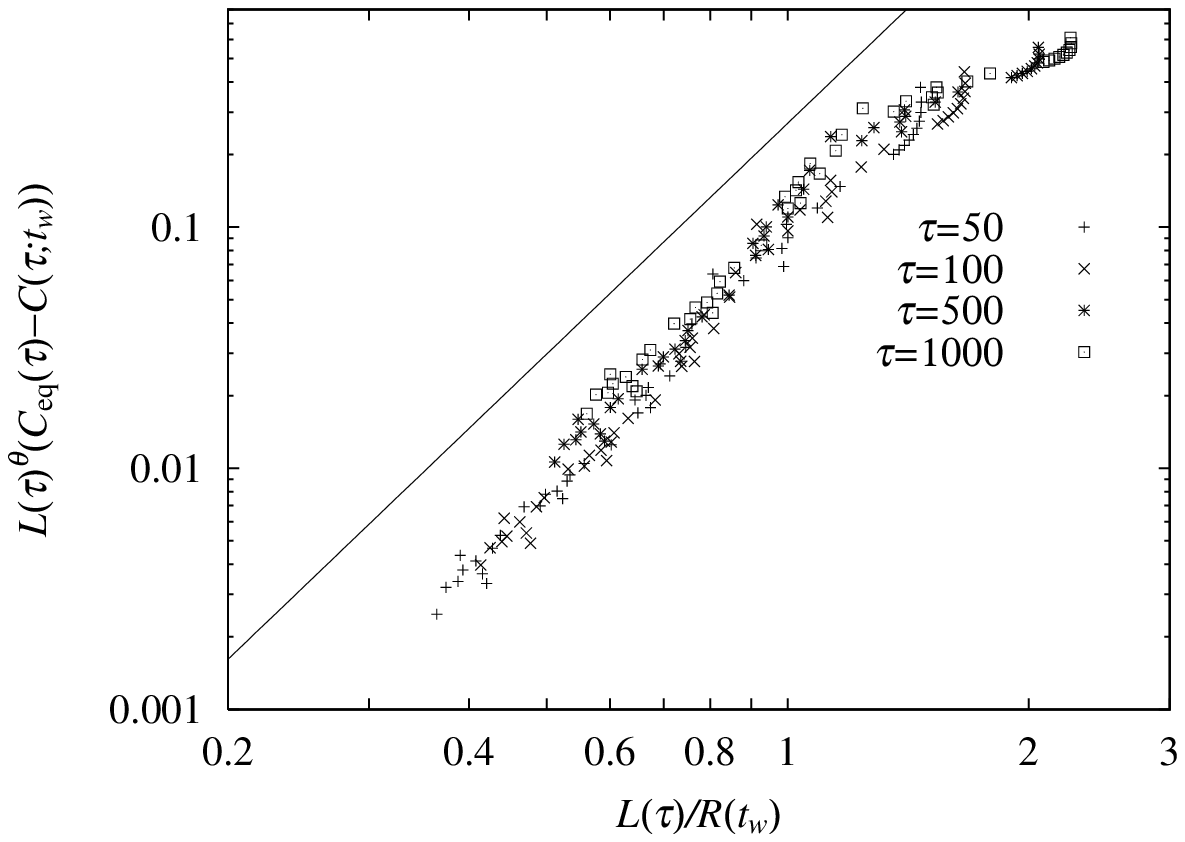}}
 \caption{Scaling plot of the correction term of $C(\tau;t_w)$. 
The line has an expected slope $d-\theta$ for small $L(\tau)/R(t_w)$
 limit. 
}
 \label{fig:coft-scal}}
\end{figure}

We observe the autocorrelation function $C(\tau;t_w)$ at $T/J=1.2$ well 
below $T_{\rm c}$ and  check the scaling  form (\ref{eqn:coft}) by making
use of the $R(t)$ estimated through $G(r,t)$ in the last section. 
According to the droplet theory, both $R(t_w)$ and $L(\tau)$ exhibit the 
same time dependence. 
Also, the microscopic length scale $L_0$ is assumed to be the same as
$R_0$. 
For fixed $\tau$, the autocorrelation function is expressed as a
function of $R(t_w)$.  
Using the estimated value of $\theta(=0.82)$,\cite{KH}  a simple linear
fitting gives  the equilibrium autocorrelation function $C_{\rm
eq}(\tau)$ in the large $t_w$ limit. 
Collecting $C_{\rm eq}(\tau)$ for each $\tau$ thus extracted, we confirm
directly the droplet prediction (\ref{eqn:ceq}) as shown in
Fig~\ref{fig:ceq} and determine the equilibrium EA order parameter.  
The value of $q_{\rm EA}$ estimated to be $0.58(1)$ is compatible with
the recent estimation from static MC simulation.\cite{Marinari99} 

Next we discuss correction to the equilibrium limit. 
The expression (\ref{eqn:coft}) suggests that the correction term
$\Delta C(\tau;t_w)=C(\tau;t_w)-C_{\rm eq}(\tau)$  multiplied by 
$L^\theta(\tau)$ becomes only a function of $L(\tau)/R(t_w)$. 
As shown in Fig.~\ref{fig:coft-scal}, we confirm this scaling
prediction.  
For the limit of $L(\tau)/R(t_w)\ll 1$, the scaling function shows 
$\left(L(\tau)/R(t_w)\right)^{d-\theta}$ 
consistent with (\ref{eqn:coft}).

\section{Discussion and Summary}

Let us compare the present results with those obtained in 3D Ising SG
models.  
In three dimensions, 
results on the growth law by numerical
simulations\cite{Kisker,Marinari98,KYT} as 
well as experiments\cite{Joh} are well fitted to a power  law as 
\begin{equation}
 R(t)\sim t^{1/z(T)},
\end{equation} 
where the exponent $1/z(T)$ is proportional to temperature $T$ and
continuously connects with the dynamical critical exponent $z$ at
$T_{\rm c}$.  
It is not clear yet how to interpret physically such a power law with
temperature-dependent  exponent $1/z(T)$. 
One of the possibilities  is the crossover observed in the preset work
on 4D Ising SG model. 
Certainly it is worth examining the crossover effect of the
critical fluctuation in the 3D model. 

On the other hand, the scaling argument in terms of the length scales
$R(t_w)$ and $L(\tau)$ successfully  explains the aging behavior of the
correlation function in the quasi-equilibrium regime  also in three
dimensions.  
Recently it has been confirmed  not only in the isothermal but also in 
temperature-shift aging processes,\cite{TYK} which are basic
experimental procedures frequently used.

To conclude, we have investigated  non-equilibrium dynamics after the
temperature quench from infinity to the SG phase in the 4D Ising
SG model using Monte Carlo simulations. 
We have studied the growth law of the mean domain size by analyzing
time evolution of the spatial replica correlation functions. 
We have found that 
the growth law shows a crossover from the critical regime to the
low-temperature one 
and its main temperature dependence within time range
of our simulation can be explained by this crossover.  

We have also analyzed the spin autocorrelation function in the
quasi-equilibrium regime. 
The off-equilibrium correction of the correlation function to its
equilibrium limit, namely, 
violation of the time translational invariance, in the quasi-equilibrium
limit can be explained by the scaling argument in terms of the
characteristic length scales $R(t_w)$ and $L(\tau)$,  
as observed already in the 3D Ising SG model.\cite{KYT2} 
These results strongly suggest that 
the analysis consisting of the two almost independent steps is promising 
for understanding the aging phenomena in low-dimensional SG systems.

\section*{Acknowledgements}
The present simulations has been performed on Fujitsu VPP-500/40 at the
Supercomputer Center, Institute for Solid State Physics, the University
of Tokyo.

\end{document}